# META RESONANT WAVEGUIDE-GRATINGS PROVIDING SELECTIVE DIFFRACTION

Guillaume Basset[1,2]

## I. ABSTRACT

The past decade has witnessed the development of a large variety of new flat optics referred to as metasurfaces [1]. These metasurfaces are relying on arrays of a large variety of phase shifting elements. This article aims at presenting a novel type of flat optics able to perform wavefront shaping and beam redirection with a high wavelength or angular selectivity, as well as an intrinsic polarization selectivity. These new elements are not relying on localized phase-shifting but on a distributed guided-mode, closely linked to Resonant Waveguide Gratings (RWG). These new optical elements are referred to as Meta Resonant Waveguide Gratings (MRWG). MRWGs have intrinsically rather low aspect ratios and can be produced with existing high-throughput manufacturing methods such as hot embossing and physical vapor deposition (PVD), making them industrially attractive. Multiple design options linked to their many degrees of freedom enable a wide range of optical properties, such as color-selective free-space combiners and diffractive couplers configurations. The properties of a simple MRWG are detailed, before unveiling a few variants to discuss the influence of some key design parameters, and providing a general design method for complex arrangements.

## II. SUPPLEMENTAL PERIODIC PERTURBATIONS IN RESONANT WAVEGUIDE-GRATINGS

Diffraction gratings and thin waveguides are two of the most widely used photonic elements. Resonant waveguide gratings (RWG) are in essence a combination of both in direct optical contact. Originally investigated in the 70s as optical couplers to planar waveguides, they have found a variety of applications in the following decades such as laser cavity mirrors, sensing devices or very resilient optical security features for the authentication of ID documents and banknotes. As a consequence of their various uses and parallel developments in the West and the USSR, they also have been called in many different ways: guided mode resonant gratings, leaky mode resonant gratings, grating slab waveguides, guided-mode resonance devices and resonant waveguide gratings. The development and applications of RWGs have been reviewed in 2017 [2]. We include a short summary of the history of their development, their use in non-zero order configuration and the path to the development of meta-RWG in annex.

Targeting the diffraction of a single band diffraction, one must realize that a compound RWG must be designed primarily with a single periodicity to couple-in mostly a single wavelength range. Any deviation from this periodic design will enable light coupling into the compound RWG at other frequencies, and therefore deviations should remain limited or be designed such as not to impact the spectral band of interest. This primary incoupling period is called the basis period.

---

[1] CSEM SA. CSEM Muttenz, Tramstrasse 99, CH-4132 Muttenz, Switzerland. gba@csem.ch
[2] Resonant Screens, gba@resonantscreens.com

Contradictorily, another spatial frequency must provide a sufficient perturbation to create the necessary interferences between the leaky-waveguided light and the incident light - or in another viewpoint to outcouple the leaky-guided mode in another diffraction order - to build up a non-waveguided diffracted beam. A good compromise between these two opposite imperatives is found to be a design of a RWG having a basis period into which periodically spaced local perturbations are introduced. This second patterning period is called the local perturbations period (LPP). The concept is illustrated below while concrete examples of such new type of RWG are provided in part III.

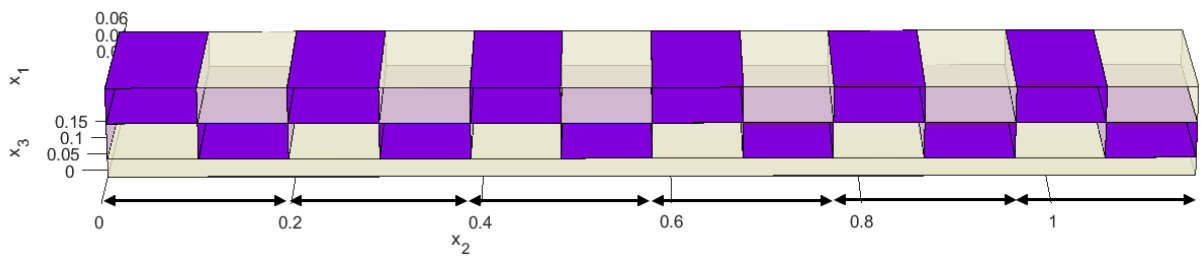

Figure 1: A schematic cross-section view of a resonant waveguide grating made with a binary grating and a double-side corrugated waveguide, for example obtained by the deposition of a thin film of dielectric on a grating. The depth of the grating equals the waveguide thickness is this specific schematic. The black arrows indicate the RWG periodicity. This periodicity is the basis period for the MRWG.

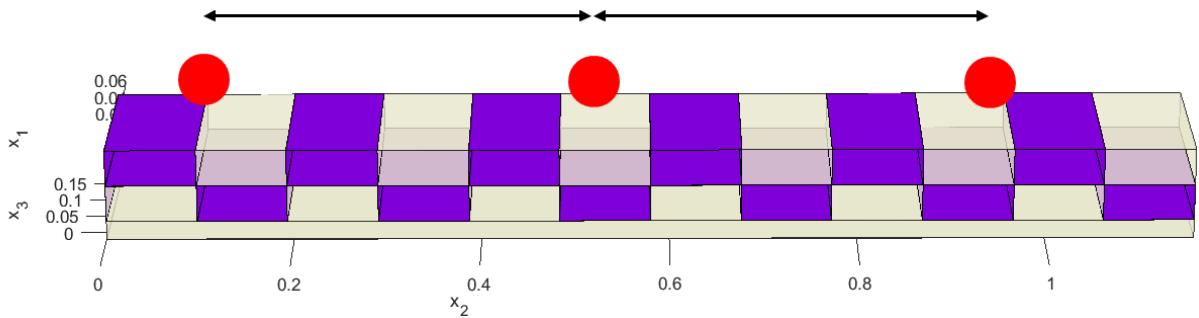

Figure 2: A schematic cross-section view of a meta-resonant waveguide-grating. The red disks indicate schematically the locations of perturbations which are introduced periodically on the RWG (no changes implemented here). The periodically space local perturbations are separated by a different periodicity, the LPP, than the basis period. The RWG is modified only locally at these locations and its structure is preserved out of these locations, to the opposite of RWG made by the superposition of multiple periodicities.

Local perturbations have a reduced impact on the light incoupling from the incidence beam, while providing sufficient light outcoupling/scattering to the waveguided light, creating periodic interferences giving rise to a diffracted beam. For binary or sinusoidal gratings, the local-perturbations patterning can for example be done on single ridges or grooves or on a few of them. Such modification of grooves and ridges is convenient to keep the manufacturing complexity low, by not impacting the waveguiding layer.

These local periodic perturbations can be of multiple types, such as modifying the profile of a groove or ridge, the height or depth of respectively a ridge or groove, its width/duty-cycle in the case of binary design, the local refractive index by for example local doping, or the addition of a nanostripe of the waveguiding material or another material etc.

The author calls such complex guided-mode resonant gratings Meta-Resonant Waveguide Gratings (MRWG). These devices are introduced with the optical characteristics of a simple MRWG configuration before describing more complex designs and a general design method.

### III. AN EXAMPLE OF A META RESONANT WAVEGUIDE GRATING

#### i. A Binary Periodic MRWG: Geometry

An example of a simple periodic and binary Meta-Resonant Waveguide Gratings is provided, for which only the duty-cycle of individual ridges and grooves are modified as an illustration purpose, before discussing other possible geometries and various design parameters in part IV and describing the general case and providing a design method in part VI. The optical behavior are here computed using the Fourier Modal Method (FMM or Rigorous Coupled Wave Analysis - RCWA) [3,4] implemented in GD-Calc in Matlab as well as experimentally measured for MRWG fabricated by UV-NIL replication and PVD.

A 1-dimensional MRWG having a periodicity of 580nm is chosen, matching its LPP, each period containing 3 grooves and ridges. Two ridges and grooves pairs have a period of 230nm followed by a single pair of ridge and groove being shorter with 120nm width. The grooves and ridges are symmetric in duty cycle providing an overall duty cycle of 50% (grooves and ridges measure 115nm or 60nm). This binary structure has a depth of 70nm made into a semi-infinite transparent polymer, here chosen as a PMMA. It is coated with an idealized full-face and deposition at normal high refractive index dielectric (Zinc Sulfide) of 40nm thickness, providing the corrugated waveguide.

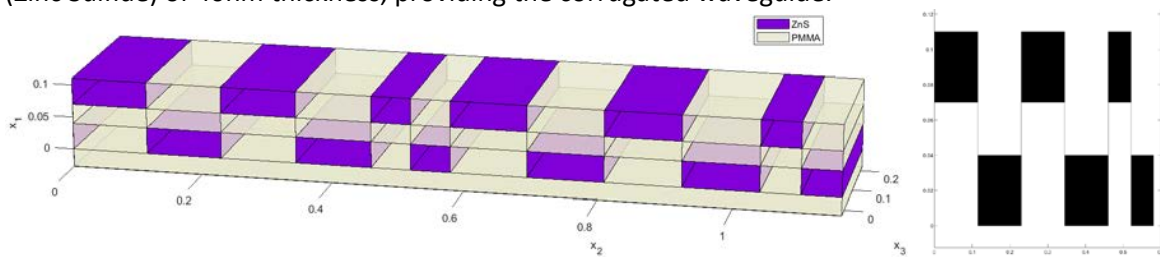

*Figure 3: Left: 3D view of the geometry of the MRWG example from GD-Calc in Matlab. Two adjacent periods are shown. The superstrate made of PMMA is not rendered. Right: Cross-section view a periodic unit cell and its dimensions. The black domain are made of Zinc Sulfide, all the white surroundings volumes are PMMA. Scale in microns.*

#### ii. A Binary Periodic MRWG: Numerical Modelling Of Optical Properties

The optical properties of this MRWG are investigated in the visible range using the RCWA toolbox GD-Calc in Matlab made by Kenneth C. Johnson. The LPP, here matching the periodicity, of 580nm is designed to diffract green light in air from grazing incidence to close to the normal and vice-versa. Below is presented the minus first order diffraction of this MRWG efficiency in both transmission and reflection as well as it's zero order transmission and zero-order /specular reflection, for TE and TM polarization for a broad range of incidence angle – the angles are defined in PMMA. The minus first order diffraction spectra for specific angles of incidences as well as the full scattering, considering all the relevant diffraction orders, again for a few selected order but only for TE polarization are included.

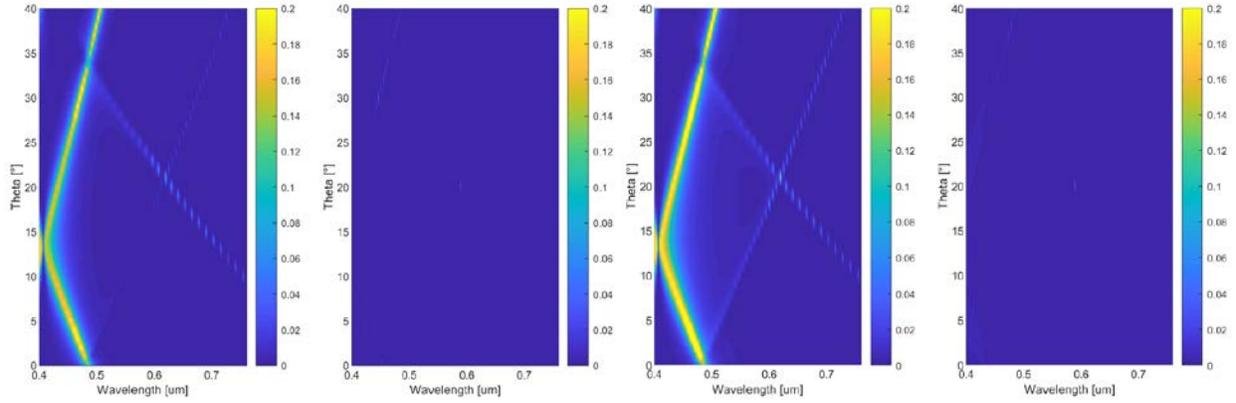

*Figure 4: 1st order diffraction efficiency for varying incidence collinear angle with light incident from the top left. Theta: Angle in PMMA of the wave incidence with respect to the plane normal. Left most: TE polarization in transmission ($T1_{TE}$), left: TM polarization in transmission ($T1_{TM}$). Right: TE polarization in reflection ($R1_{TE}$), right most: TM polarization in reflection ($R1_{TM}$).*

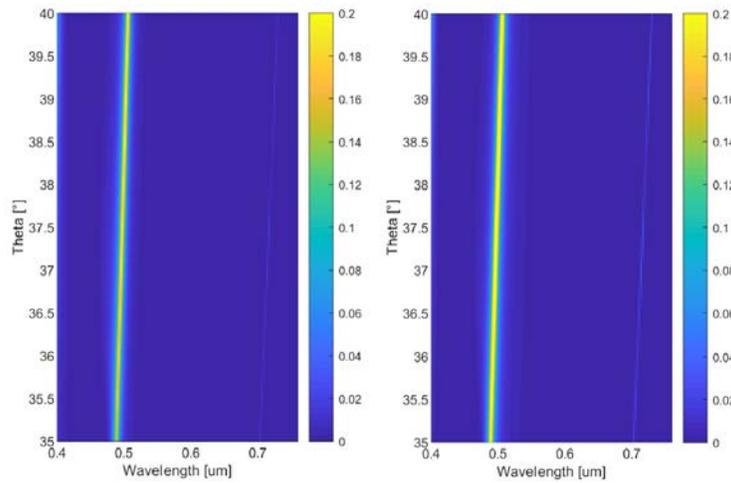

*Figure 5: Zoom on the minus first order diffraction for grazing collinear incidence between 35 and 40° zenithal angle in PMMA, only TE. Left: Transmission. Right: Reflection*

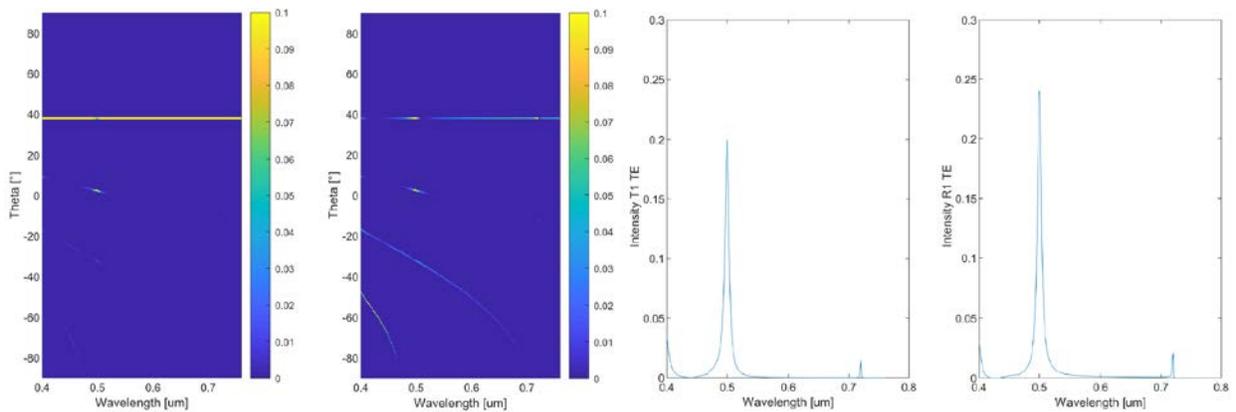

*Figure 6: Full transmission and reflection efficiency at 38° from normal collinear incidence (from the left side) in PMMA in TE polarization. Left most: Transmission, left: Reflection. Minus first order diffraction spectrum TE polarization. Right: Transmission, right most: Reflection.*

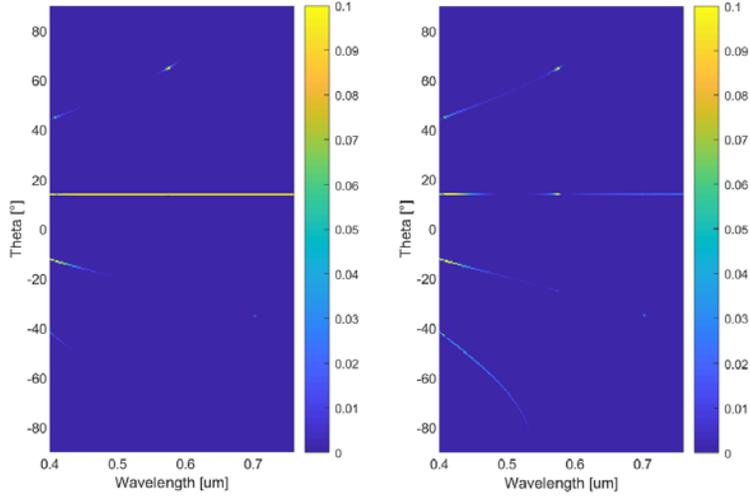

*Figure 7: Full transmission and reflection at a moderate oblique angle of 14° collinear incidence from normal (incidence from top left) in PMMA in TE polarization. Left: Transmission, Right: Reflection.*

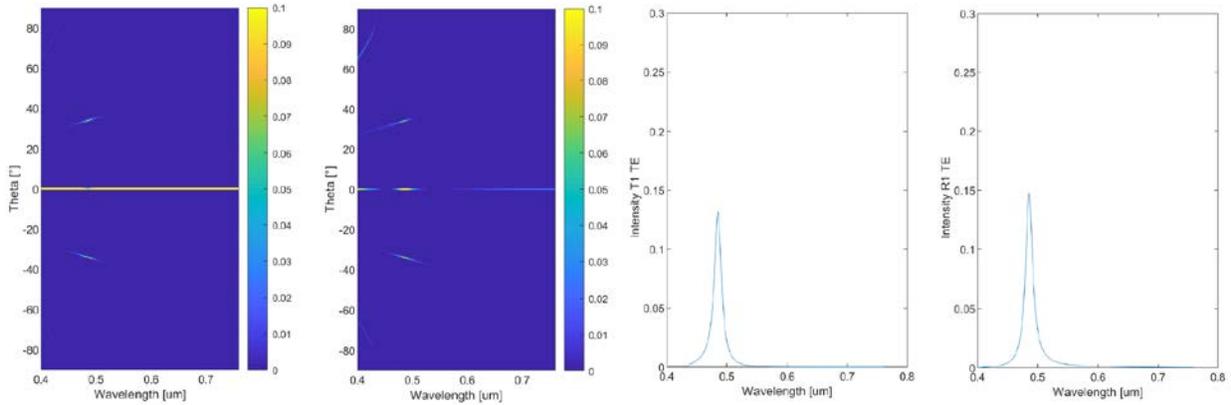

*Figure 8: Full transmission and reflection at normal incidence for TE polarization. Left most: Transmission, left: Reflection. Minus first order diffraction spectrum at normal incidence for TE polarization. Right: Transmission, right most: Reflection.*

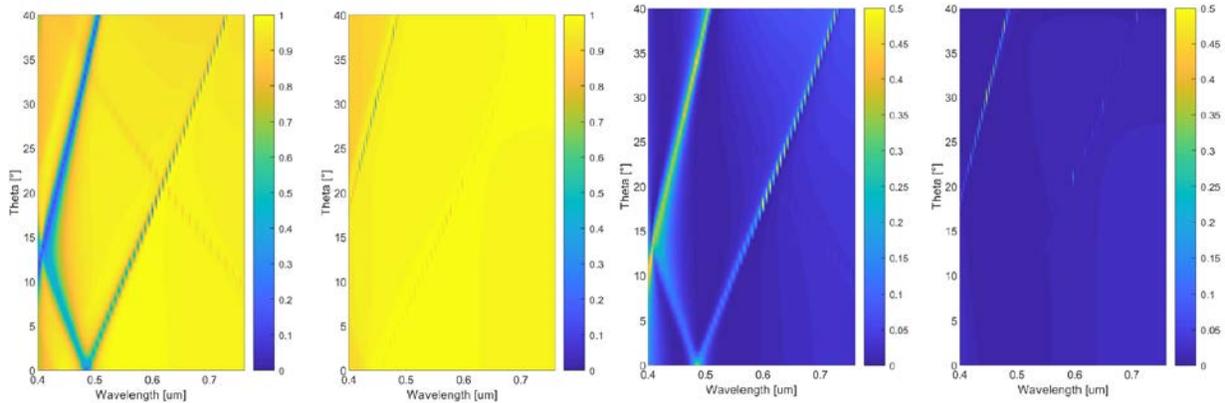

*Figure 9: Zero-order / direct transmission (left) and reflection (right) for varying collinear incidence angle in PMMA. Leftmost: TE polarization transmission, left: TM polarization in transmission, right, TE polarization in reflection, and right most: TM polarization in reflection.*

Numerical modelling predicts that this MRWG diffracts only violet, blue or green light depending on the incidence angle for TE polarized light while diffracting significantly no other wavelengths of the visible range in these broad range of angles of incidence. In grazing incidence angle – 38° in PMMA is around 70° off-normal in air – it will diffract blue/green light to close to normal incidence and vice-versa. This while maintaining a very high transparency for both polarization across the full wavelength range – except close to these resonances. This optical behavior is distinct from classical RWG and to our knowledge not reported so far.

### iii. Experimental Results

Experimental spectroscopy results on a fabricated samples are presented. A sample is fabricated by UV-NIL replication into a UV-curable solgel (called here UV-polymer, n=~1.54 at 510nm) from an e-beam photoresist master. The replicated structure is coated with the high refractive index dielectric (Zinc Sulfide) by PVD before being embedded by overcasting the UV polymer and providing a cover glass.

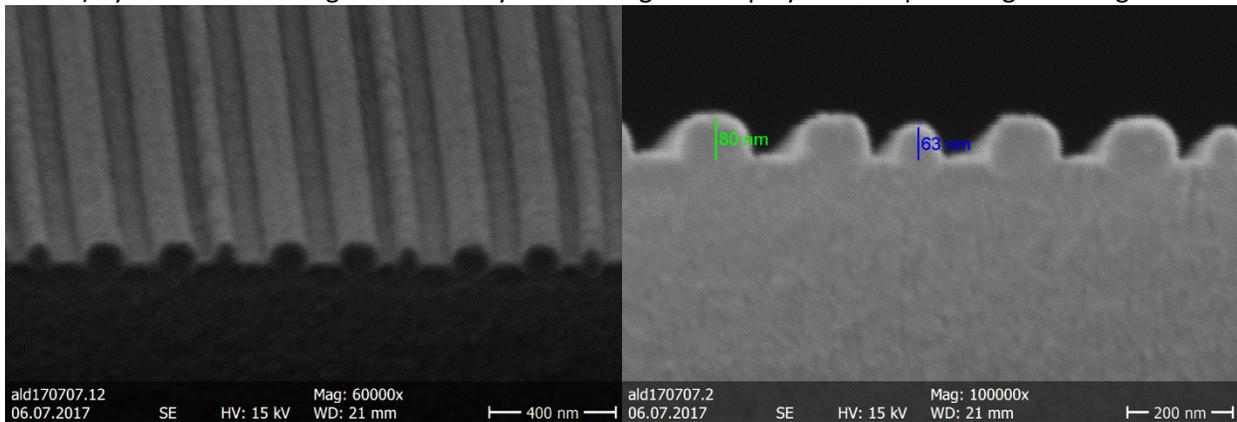

*Figure 10: SEM image of a UV-polymer replica of a master made by e-beam lithography of the MRWG pattern describe above. Left, tilted top view. Right: Cross-section view. The deposited waveguide thickness is measured at 42nm of Zinc Sulfide (ZnS, not shown here).*

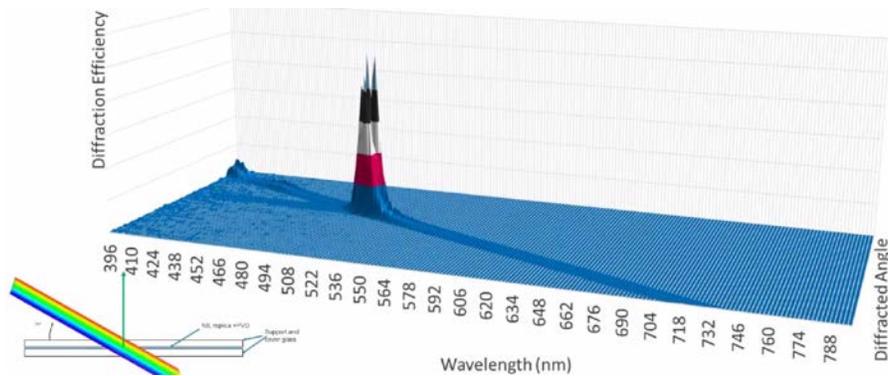

*Figure 11: Measured -1$^{st}$ diffraction order in TE polarization for a 70° off-normal collinear incidence with a Spectrascan PR730 in a goniometric setup. The two peaks are not observed by eye or photograph and are considered an artefact of the instrument.*

Additionally, spectroscopy is made with a Perkin Elmer Lambda 1050 equipped with a goniometer for different incidence angle. A polarizer is selecting the TE polarization and a slit provides a narrow illumination beam (unmeasured beam divergence).

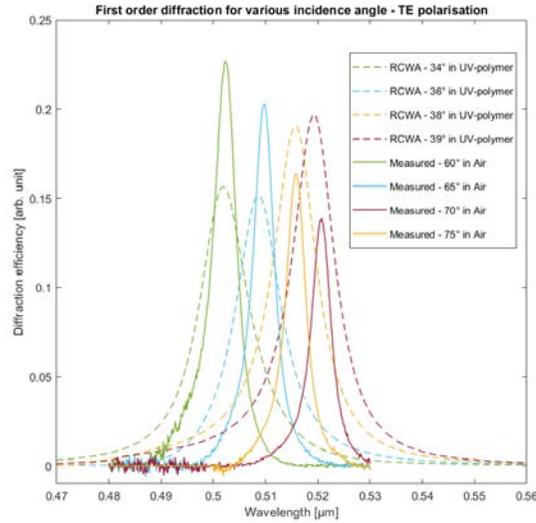

*Figure 12: Measured (solid lines) and RCWA-computed (dash) minus first order diffraction spectra in TE polarization of the MRWG described above (70nm depth, 51nm ZnS deposited) for different angles from the normal to the grating plane (60, 65, 70 and 75° in air, 34, 36, 38 and 39° in the UV-polymer) of collinear incidence. The efficiency is not measured as the impeding beam was larger than the MRWG at the grazing incidence, explaining partially the lower measured efficiency for more grazing angles.*

### iv. Physical Analysis Using the Grating Theory and Guided Mode Effective Index

Using a slab waveguide approximation, the fundamental mode for TE polarization ($TE_0$) of a 40nm ZnS waveguide in PMMA is computed at Neff = 1.676. For an incidence angle of 38° from the grating plane (collinear configuration) in PMMA, minus first order diffraction for a grating period of 193.3nm (580/3) into the $TE_0$ mode is possible for a wavelength of 502nm - equivalent to a minus third order diffraction for the 580nm period grating.

For the reciprocal case, with an incidence angle of 2.1° to the grating plane in PMMA, the resonant incoupling into the corrugated waveguide $TE_0$ mode (Neff = 1.676) can occur for a 501.9nm wavelength light with the minus first order diffraction considering a period of 290nm (580/2) - equivalent to a minus second order diffraction for the 580nm period grating.

The effective index of a corrugated waveguide is different from the one of a slab waveguide, in addition to being leaky. Slab waveguide approximation, if providing good agreement for shallow corrugation, will exhibiting discrepancies to numerical modelling and experimental results, especially for deep corrugation. Despite this limitation, this elementary analysis matches well the numerical modelling and the experimental results to predict possible resonances.

The variation of the wavelength of the minus first order diffraction in PMMA of this MRWG over variable angles of incidence is matching the zero-order resonant reflection and is the resonance of a simple RWG considering the fundamental mode index and the grating theory with the basis period. The minus first order diffraction in PMMA to 2.1° from normal in PMMA can be explained with the grating theory considering the 580nm of the LPP.

However remains the questions of why in this design the second and third diffraction orders are efficient to couple light into a guided mode, without providing broadband diffraction in free-space, and why first order diffraction is limited to the resonantly incoupled light around 502nm wavelength – resulting in a wavelength selective minus first order free-space diffraction.

It is noted that the impact of the local perturbations is different, on one side for the diffractive coupling of the incident beam into the corrugated waveguide whose behavior is quasi the one of the basis period, and on the other side for the significant non-guided diffraction of the incident field experiencing the LPP. A tentative explanation is proposed and is linked to multiple factors. Firstly the nanostructure dimensions allow only a few diffractive orders for the shortest pseudo-periodicity, the basis period. It only allows the coupling into the waveguide fundamental mode (in addition to zero-order diffraction). This is why individual interferences created by individual grooves and ridges are almost only incoupling light into the corrugated waveguide. Additionally, guided-mode confinement and field accumulation in the corrugated waveguide allow an electromagnetic field build-up having a periodic modulation with the LPP and each local perturbation acting as a scattering center for the guided mode. This allow the incident beam to interfere with the guided mode out of the zero-order diffraction.

In other words, the local periodic perturbations are small enough not to create significant direct diffraction by phase shifting of the incident field, not to impact significantly the incoupling into the guided-mode nor to provide direct broadband diffraction, but are sufficient to scatter out periodically a fraction of the guided mode, producing a periodically modulating guided mode, which give rise to the diffraction of the incident light matching the guided-mode frequency. This makes possible a wavelength selective resonant diffraction while having a very high transparency to all non-resonant wavelengths - which are experiencing negligible phase shifts when passing through the MRWG. The Fourier Modal Method modelling and the experimental results are compared with time-resolved modelling using Final Difference Time Domain (FDTD, Omnisim software from Photon Design).

## v. Time Resolved Modelling with FDTD

A two-dimensional modeling is performed using periodic boundaries. Multiple MRWG periods are designed and computed in order to have a 2 Pi phase difference between the two sides of the cells for the excitation source which is not parallel to the nanostructure plane. In this examples, for an angle of incidence of three degrees for a central wavelength of 470nm in a medium refractive index material (n≈1.54), ten periods (a cell size of 5.8 microns) provides a good phase continuity of the excitation light:

Phase shift = tan (excitation angle) x cell size
        = tan (3°) x 5800nm
        ≈ 304nm

The effective wavelength of the excitation for 470nm light is close to 305nm (n=1.54). For excitation wavelength away from 470nm, the phase shifts at the periodic boundary of the cell is moving away from 2 Pi, here by less than +/- 10%. Given the limited resolution of the far-field extraction of this FDTD software in these configurations, we observe qualitatively the field propagation, diffraction and localization for different excitation wavelengths.

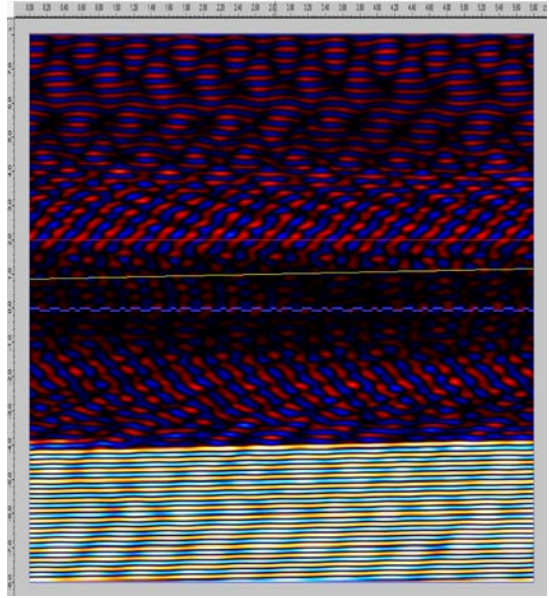

*Figure 13: Cross-Section of the transverse electric field computed in Omnisim of a 2D model of ten periods of binary MRWG (high refractive index of the nanostructures in blue) after being excited by a 30fs pulse of 440nm wavelength (excitor in yellow, tilted by 3degree with respect to the nanostructure plane) .The transmitted light is visible in the bottom third. The scale is in micron and the scale is stretched on the horizontal axis due to the computing configuration.*

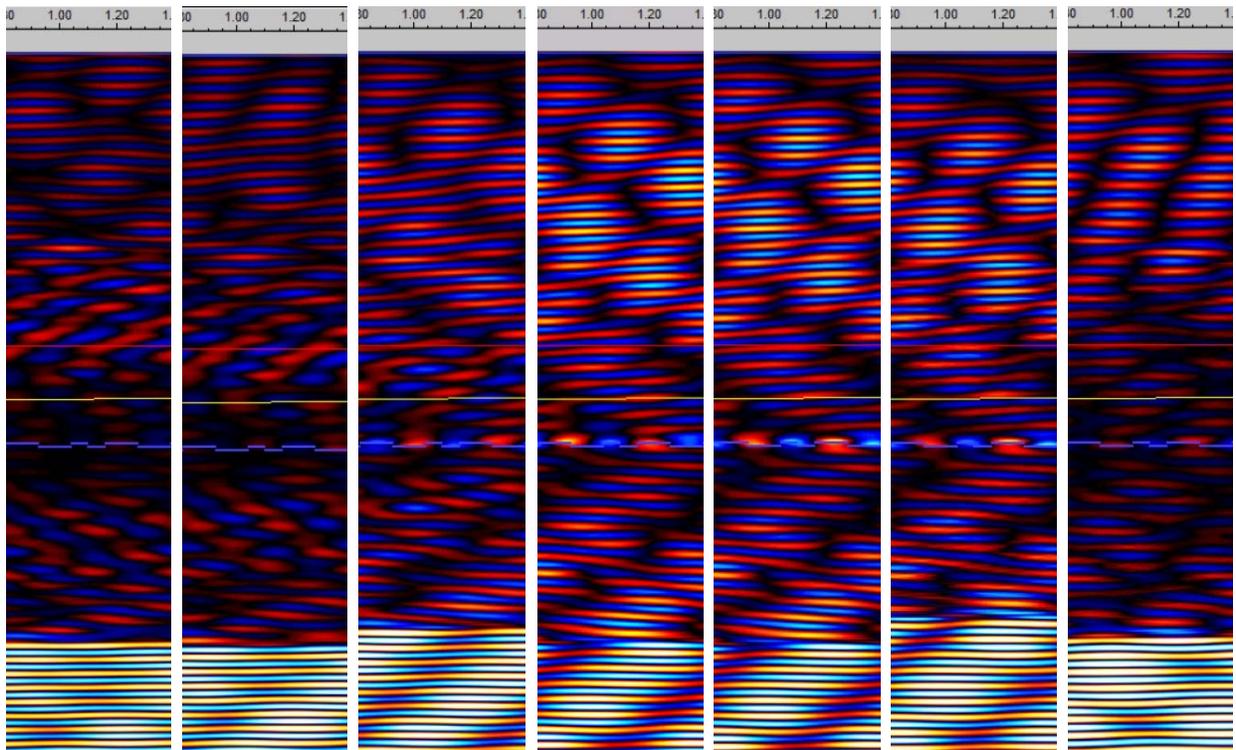

*Figure 14: Cross-section view with intensity of the transverse electric field computed in Omnisim of a 2D model of ten MRWG periods (high refractive index of the nanostructures in blue) after being excited by a pulse of 30fs (excitor in yellow) after light propagation in the MRWG. The view is cropped to one MRWG period only (580nm). From left to right, excitation wavelengths of 440nm, 450nm, 460nm, 470nm, 475nm, 480nm and 490nm. The periodicity of the computation cell allows phase continuity of the excitation light on the left/right borders but do not maintains field continuity in general for non-zero diffracted orders.*

The diffracted light, corresponding to first order diffraction of the 580nm periodicity is well visible in this computation, both for reflected and transmitted light. The wavelength selectivity is quite narrowband with peak diffraction efficiency around 470 – 475nm wavelength and much lower diffraction efficiency at 450nm and at 490nm. This matches qualitatively the results of Fourier Modal Method simulations for a 3 degree incidence angle as can be seen in figure 4. The time-resolved simulation additionally shows (not visible in Figure 14) that the free space diffraction appears with the guided mode building up. The diffraction efficiency is very low for the first impeding wavefronts but increases drastically once sufficient light is incoupled in the monomode waveguide. This confirms that the wavelength and angular selective incoupling in the waveguide induces the wavelength and angular selectivity of the MRWG in the far field.

vi. An Example of Sinusoidal Profile MRWG

To further illustrate this optical properties and reinforce the validity of the analysis with the grating theory and guided mode computation, a numerical study of the equivalent MRWG with sinusoidal profiles is done. The depth of the structure is increased to 80nm to get similar modulation and efficiency to a 70nm deep binary grating. The sinus profiles are approximated by staircases in 10 equally thick layers of 8nm. The ZnS top profile is identical to the under profile. Note that the narrow groove profile is not exactly equivalent to the binary grating geometry described above but this design is similar in the way that one period out of 3 is narrow down to 120nm while the two adjacent periods are 230nm.

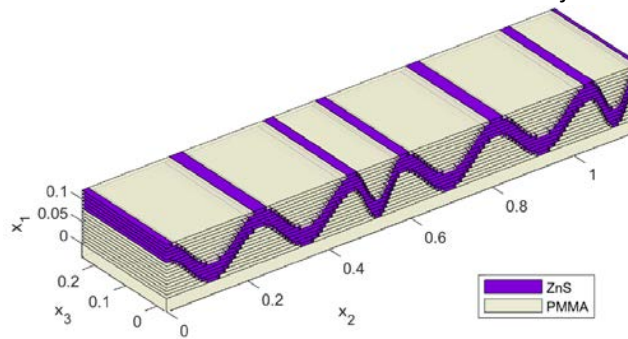

Figure 15: 3D view of the geometry of the staircase approximated geometry of a sinusoidal MRWG from GD-Calc in Matlab. Two adjacent periods are shown. The superstrate made of PMMA is not rendered. Scale in microns.

RCWA is used to compute the diffraction of this staircase approximated sinusoidal profiles MRWG.

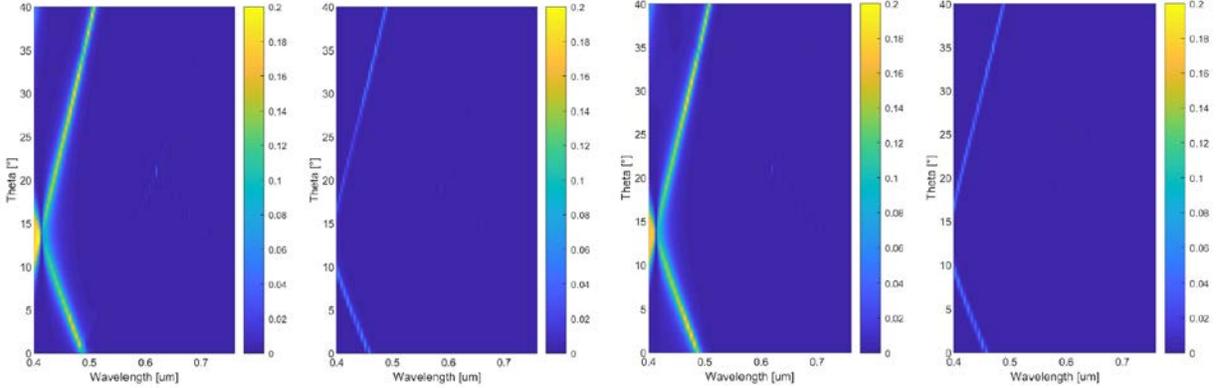

*Figure 16: Minus first order diffraction for varying collinear incidence angle between 0° and 40° from normal in PMMA, sinus MRWG approximation. Left most: TE polarization in transmission, left: TM polarization in transmission, right, TE polarization in reflection, right most, TM polarization in reflection.*

An overall similar behavior is observed with the binary MRWG, although with a lower diffraction efficiency in TE polarization and a higher efficiency in TM polarization. This points to the key role of the waveguiding-mode resonance and not to local or geometry-specific resonances. The nature of the local perturbations impacts the behavior of the MRWG and various properties can be obtained with the many degrees of freedom of MRWG designs. Some examples are provided in part IV.

## IV. INFLUENCES OF SOME DESIGN PARAMETERS

### i. Various Periodic Perturbations

Beyond a specific example, the variety of possible designs of MRWG is illustrated by providing examples of various periodically-spaced local-perturbations, and the impact on the optical performances obtained. The variations are derived from the previous examples in order to have a comparison basis. The results are computed using RCWA modelling except experimental results which are noted as such.

MRWG can be designed using a narrower small pair or ridge and groove constituting the local perturbations. An example is provided having the same depth (70nm), High Refractive Index (HRI) thickness (40nm) and the same periodicity than the reference design provided in part III. However the narrow grooves and ridges are reduced to 40nm, while the two other periods are increased to 250nm.

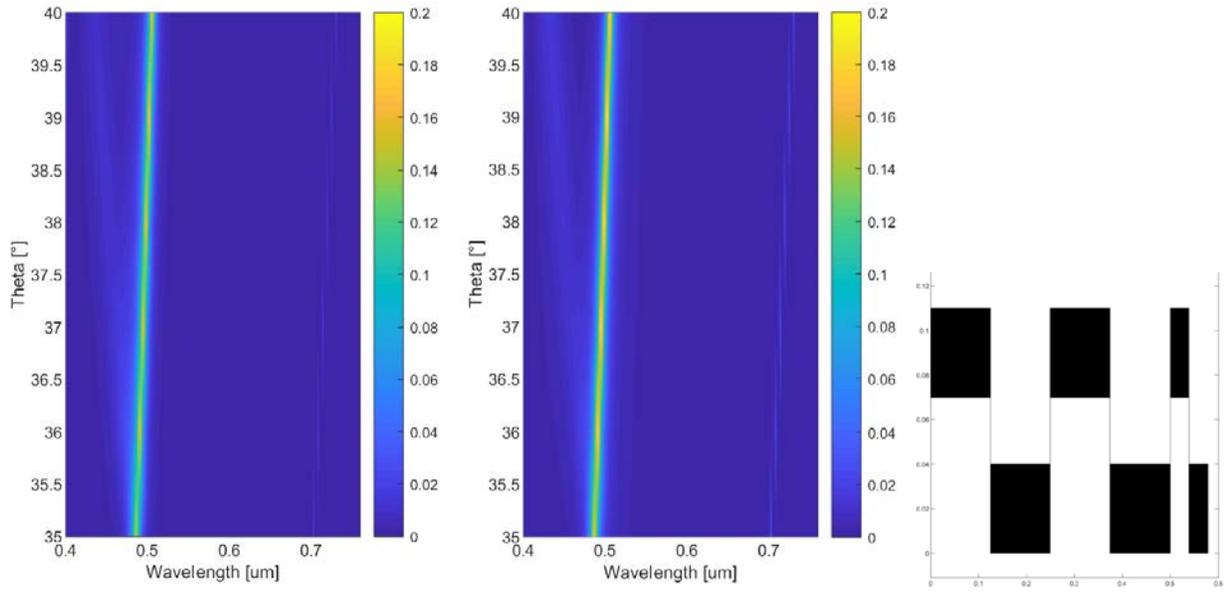

*Figure 17: Minus first order diffraction for grazing collinear incidence between 35 and 40° zenithal angle in PMMA, only TE. Smaller period as a local perturbation. Left: Transmission. Right: Reflection*

The diffraction remains similar with slightly more diffraction in the violet and blue range for grazing incidence angles (in air). This shows another example of the relative stability of the optical performances to significant geometrical variations.

MRWG can be designed as well using a single narrower ridge (or groove) than its counterparts. An example with only one narrower ridge with a 60nm width is provided. The other two ridges and three grooves are set equal, at 104nm to keep the overall periodicity at 580nm. The binary structures have the same depth (70nm) as the reference design and a thicker HRI coating (60nm).

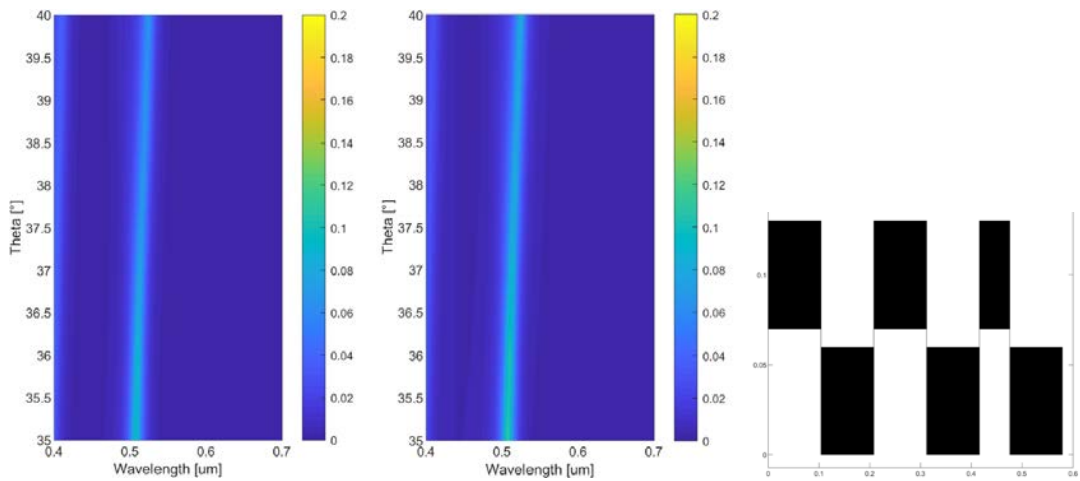

*Figure 18: Minus first order diffraction for collinear grazing incidence between 35 and 40° zenithal angle in PMMA, only TE. Only one ridge as a local perturbation. Left: Transmission. Right: Reflection*

The diffraction bandwidth is spread to roughly 20nm FHWM while the efficiency is lower. The reflection efficiency is slightly higher than the transmission one. This can be inverted by having a narrow-groove instead of a narrow-ridge (not shown here) with an incidence from the top.

MRWG can be designed with equal width for all ridges and grooves by modulating the ridge height and/or groove depth. The nanostructures are not binary anymore, but in this case do not require thin line patterning. An example is provided with a single higher ridge, at 110nm, the height of two other ridges are kept at 70nm as the HRI thickness at 40nm. All ridges and grooves have equal width ~97nm.

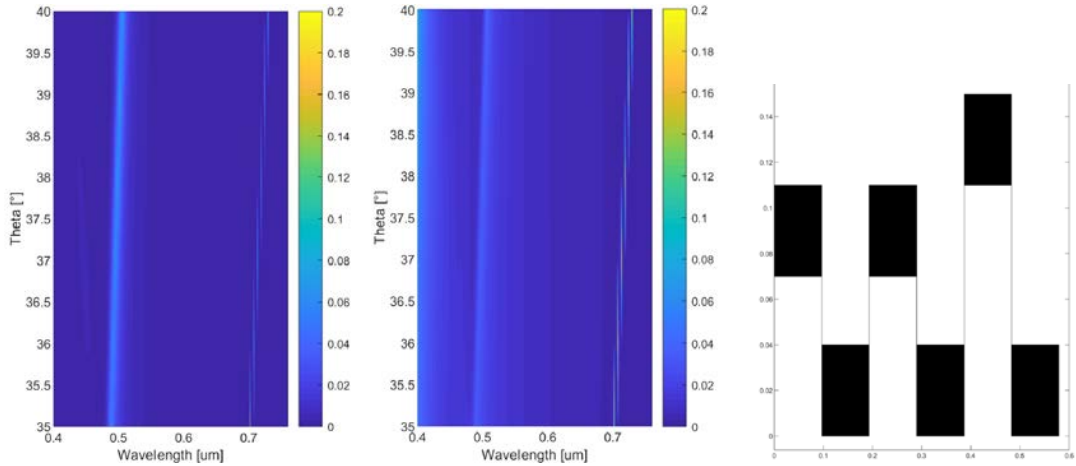

*Figure 19: Minus first order diffraction for grazing collinear incidence between 35 and 40° zenithal angle in PMMA for ~97nm ridges and grooves, coated with 40nm ZnS. 70nm ridge height and every three ridge, 110nm height. Left: Transmission. Right: Reflection*

One can observe than while the diffraction remains mostly monochromatic in transmission, it becomes quite broadband in reflection, behaving more similarly to a classical 580nm period grating.

Inversely, MRWG can be designed with one ridge height or groove depth being lowered. An example is provided with a single lowered ridge, at 30nm, the height of two other ridges are kept at 70nm above the grooves and the HRI thickness at 40nm. All ridges and grooves have equal width ~97 nm.

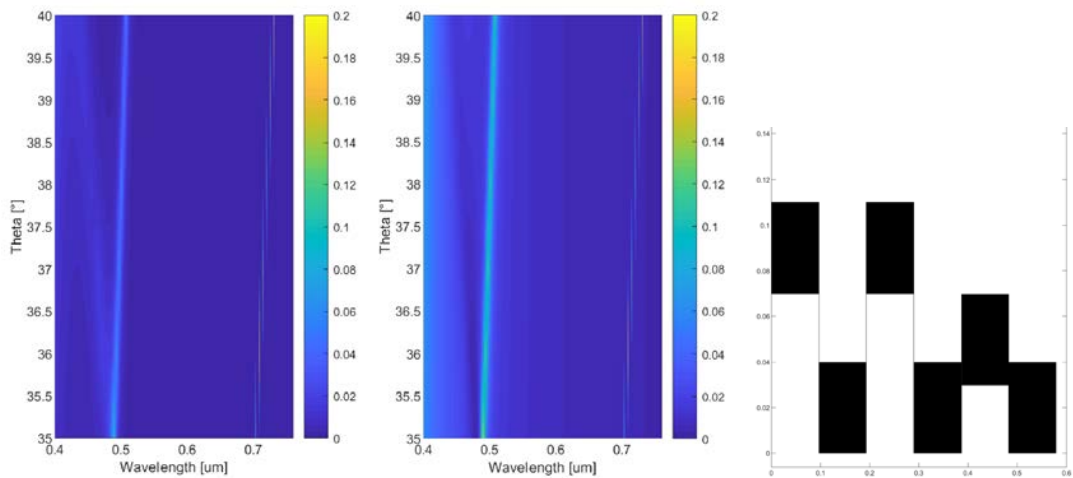

*Figure 20: Minus first order diffraction for grazing collinear incidence between 35 and 40° zenithal angle in PMMA for ~97nm ridges and grooves, coated with 40nm ZnS. 70nm ridge height and every three ridge, 30nm height. Left: Transmission. Right: Reflection*

In this case again, the diffraction efficiency is lowered compared to local perturbations made by ridge/groove width modulation. Here again a stronger transmission/reflection dissymmetry is observed.

## ii. Quality Factor and MRWG Depth

Various design parameters of MRWGs influence the bandwidth of the minus first order diffraction. Among those, the nanostructures' depth is a key parameter. Overall, a less corrugated waveguide enables longer average waveguiding path length, translating in a more wavelength-selective resonance. Such shallower nanostructures exhibit smaller diffraction bandwidth in the zeroth and the minus first diffraction order. This is illustrated by computing a shallower version of the binary MRWG described in part III with a depth reduced from 70nm to 50nm. The HRI thickness is increased to 62nm to compensate partially the change of the coupled mode index. The structure has been realized and characterized.

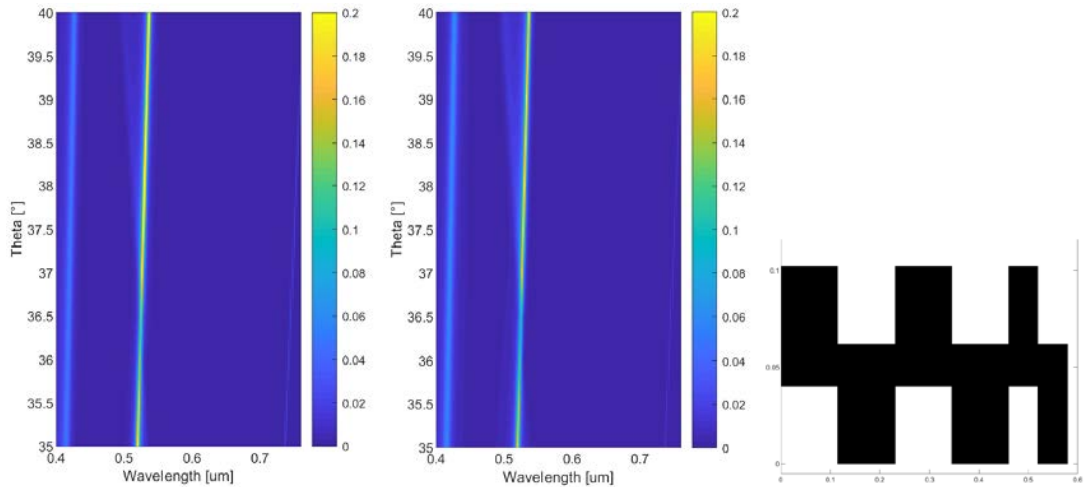

*Figure 21: Minus first order diffraction for grazing collinear incidence between 35 and 40° zenithal angle in Ormocomp, only TE. The HRI thickness remains was increased to 62nm, the grooves and ridges width are kept at 115nm and 60nm but the binary structure is shallower, with a depth of 40nm. Left: Transmission. Right: Reflection*

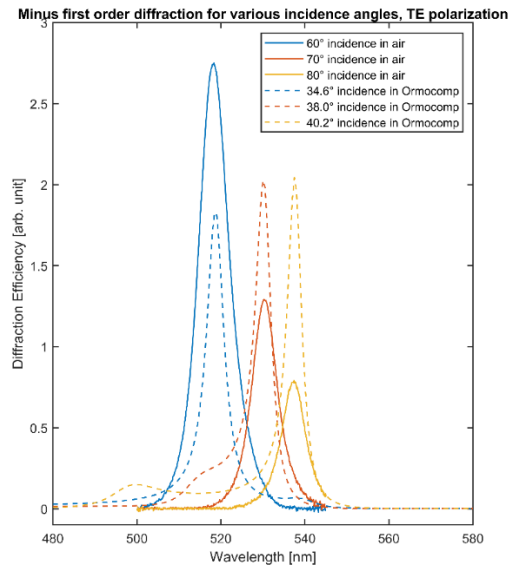

*Figure 22: Measured (solid lines) and RCWA-computed (dash) minus first diffraction order for a shallower binary MRWG design. Comparison between measurements at 3 different grazing collinear incidence in air and RCWA computation (angles in Ormocomp). The lower efficiencies measured for more grazing incidence can be partially explained by the limited size of the sample, smaller than the beam cross-section, providing a not quantitative measurement. The FWHM bandwidth is measured*

*slightly below 10nm. The measurement beam of the Perkin Elmer Lambda 1050 is not perfectly collimated (un-measured angular distribution), which is widening slightly (not estimated) the diffraction bandwidth.*

The quality factor of the first order resonance can be estimated to be above 50 in this experimental result. Using high quality dielectric films, in which centimeter-scale propagation is experimentally obtained [5], very long propagation-length resonances could be obtained. Theoretically very high quality factor are reachable, however they will be limited by other parameters such as the beam dimension and coherence in most practical cases.

### iii. Polarization and Waveguide Thickness

Very thin dielectric waveguide can only guide the fundamental TE and TM modes. No cut-off with respect to the dielectric thickness is existing and extremely thin waveguides can be realized, however $TE_0$ and $TM_0$ modes have different effective indexes – as well as higher waveguide modes. This naturally splits the resonances and diffraction of TE and TM polarization in most cases, for a given angle of incidence, to different wavelengths.

Using the slab waveguide approximation, the 40nm ZnS waveguide has for a wavelength of 500nm effective indexes of 1.708 for $TE_0$ and 1.572 for $TM_0$ when embedded in a 1.53 refractive index material. The $TM_0$ guiding is therefore much weaker and larger basis period are required to couple light into this mode.

An example of a larger basis period MRWG with a thicker bimode waveguide (190nm ZnS) is discussed. The $TE_0$ and $TM_0$ effective indexes are in this case estimated at 2.228 and 2.144 at this wavelength with the same approximation. The MRWG is binary with a basis period of 233nm, a LPP of 700nm and a depth of 30nm. The fundamental $TM_0$ is exited with a TM incident light.

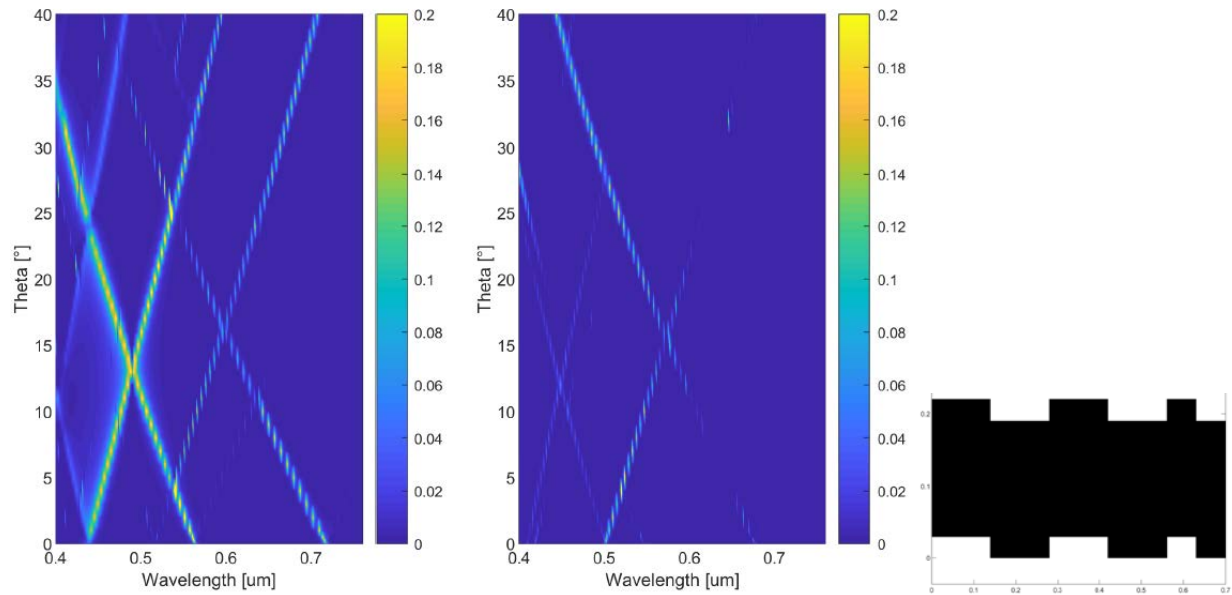

*Figure 23: Minus first order diffraction in transmission in collinear incidence, vertical axis angle in Ormocomp, with left TE polarization, center TM polarization. Right: Cross-section schematics of the MRWG with the ZnS represented in black and all the immediate surroundings being Ormocomp.*

The first order diffraction by the LPP in Ormocomp in TE polarization at normal incidence at λ ~430nm and λ ~560nm corresponds to the excitation of the $TE_1$ mode by respectively the third diffractive order of the LPP (first order diffraction by the basis period of 233nm) and the second diffractive order of the LPP. The effective index of the first TE mode is estimated at 1.834 and 1.629 for these two wavelengths. The diffraction resulting from the excitation of the fundamental $TE_0$ mode for λ ~715nm is much weaker than the one resulting from excitation of $TE_1$.

On the opposite, for incidence in TM polarization, first order diffraction by the LPP in Ormocomp is the strongest at λ~=505nm for normal incidence, and corresponds to $TM_0$ excitation by the third diffractive order of the LPP (first order diffraction by the basis period of 233nm).

This MRWG exhibits the possible selective diffraction in both TE and TM polarization, with a clear wavelength split for the two polarizations due to the large different of waveguide mode index. The computed non-waveguided diffraction are relying mostly on the first waveguide mode in TE polarization while on the fundamental mode in TM.

It has been demonstrated that thicker waveguides having a high enough index contrast with respect to their substrate can enable the coalescence of two diffractive orders to obtain broadband reflection in RWG [6]. This approach could be investigated as well in MRWG in the infrared wavelength range.

    iv.      Red, Green and Blue Selective, High Transparency and Low Dispersion Combiner

In order to benefit from the wavelength selectivity and from the high transparency, 3 MRWGs are designed to diffract selectively a blue, green and red spectral band from 70° degree off-normal to normal incidence. The 3 MRWGs are successively replicated on a glass plate by UV-NIL and coated with ZnS by a PVD. Each ZnS coated nanostructures are fully embedded by the UV-NIL of the successive MRWG replication. The 3[rd] MRWG is embedded by a last UV-NIL and an encapsulation glass, resulting with a stack of 3 MRWG located between two 1mm thick glass plates. The diffraction of this glass encapsulated stack is measured with the Spectrascan goniometric setup described above.

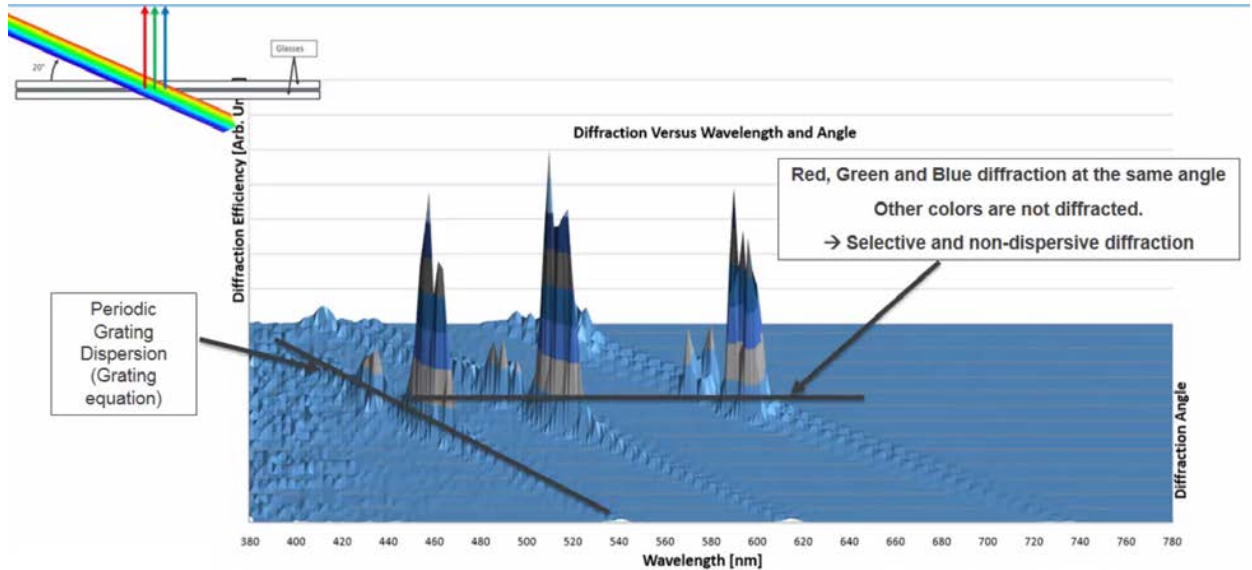

*Figure 24: Free-space diffraction spectrum measured in TE polarization for a stack of 3 MRWG fabricated by successive UV-NIL replication on a borofloat glass and PVD coating with ZnS. The 3 MRWG have binary design, their masters were fabricated ny e-beam lithography and contact copied ny NIL to fabricate NIL stamp. The light source used is a fiber coupled halogen bulb, collimated close to infinite focus with a fiber collimator and hence has a low temporal coherence.*

Despite a lower efficiency, MRWGs compares favorably to meta-surface multi-wavelength achromatic combiners [7] in terms of transparency and absence of perturbations to other wavelengths, which is highly relevant when high broadband transparency at multiple angles is needed. The stacking of three MRWG allows a large tunability as each MRWG can be optimized fully independently. The fabrication complexity is moderate except for the e-beam lithography mastering, as the three UV-NIL replications and PVDs were done in less than a day with widely available laboratory equipment.

v. An Example of a MRWG Coupler

MRWG can be designed to diffract light from/into lightguides as they can diffract from/into free-space. An example of a MRWG is provided, designed to diffract a portion of the visible light range in Total Internal Reflection (TIR) in a glass lightguide, with light incident around normal incidence.

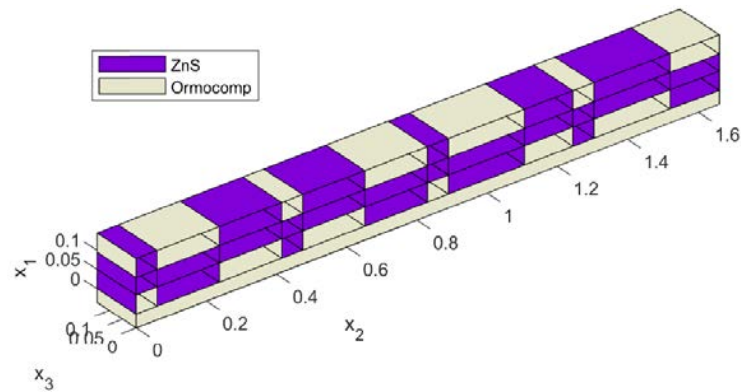

*Figure 25: The geometry of a ZnS coated MRWG embedded in Ormocomp is provided as a sketch. The MRWG period consist of 6 basis periods of 277nm with LPP of 415nm made by single narrow ridges or grooves. The nanostructures are binary with a depth of 50nm and the waveguide layer consist of 95nm of ZnS.*

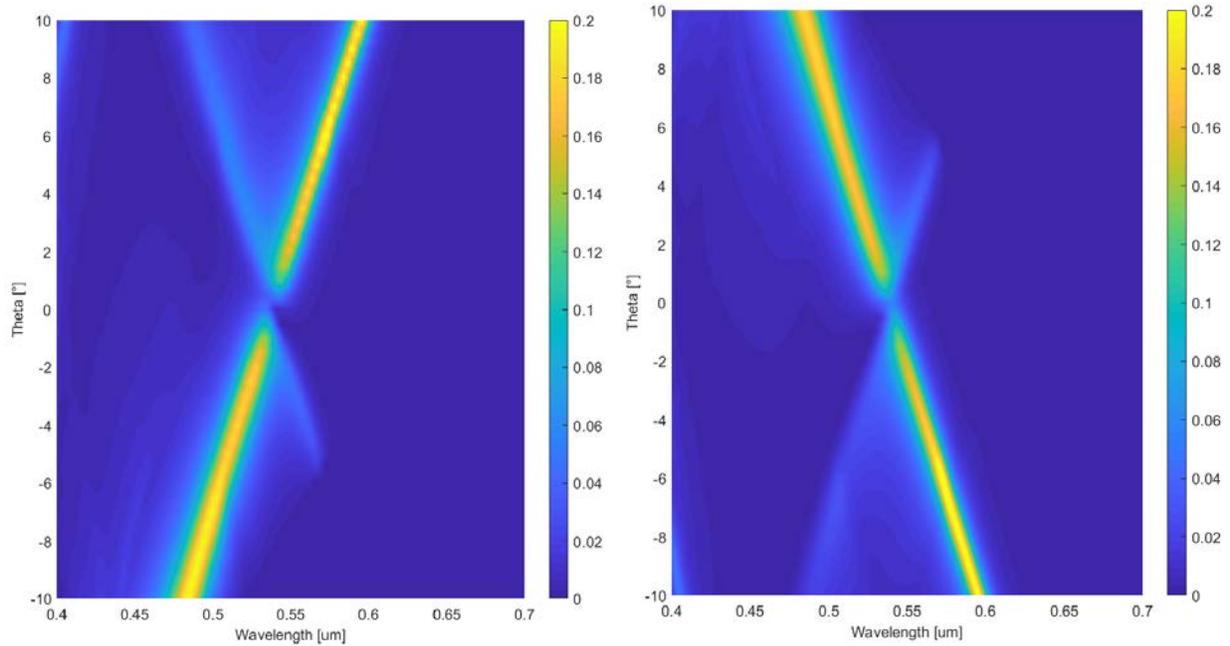

*Figure 26: RCWA computed -4th diffraction order (left) and 4th diffraction order (right) in transmission for TE polarization for varying collinear incidence angle from the normal in Ormocomp. The -4th and 4th diffraction order are diffracted to guided propagation in the Ormocomp in TIR with respect to an air interface, making this structure a color selective but not color stable light coupler. The diffraction is significantly less efficient at normal incidence due to a relative symmetry of the structure, likely creating a competition between counter-propagating guided-modes having the same frequencies.*

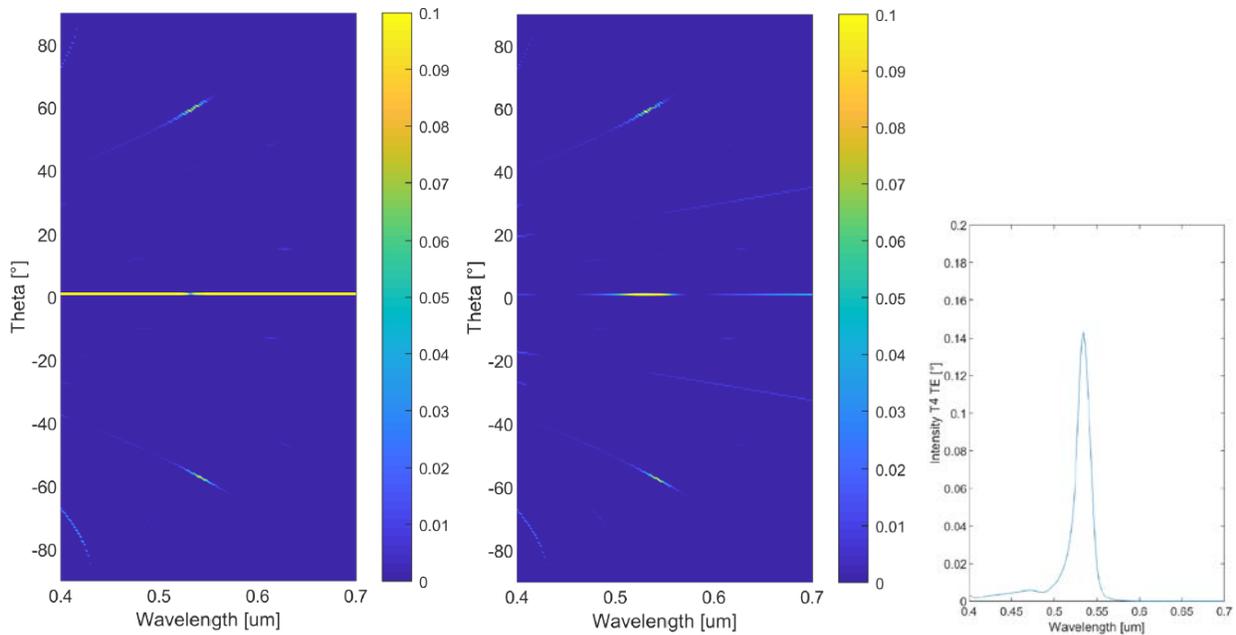

*Figure 27: Light scattering in TE polarization in transmission (left) and reflection (center) for light incident at 1° from normal in collinear incidence, angles in Ormocomp. The -4th and 4th diffraction order can be observed and correspond to guided propagation in the Ormocomp in TIR with respect to an air interface. Right: 4th diffraction order spectrum at this 1° off-normal collinear incidence.*

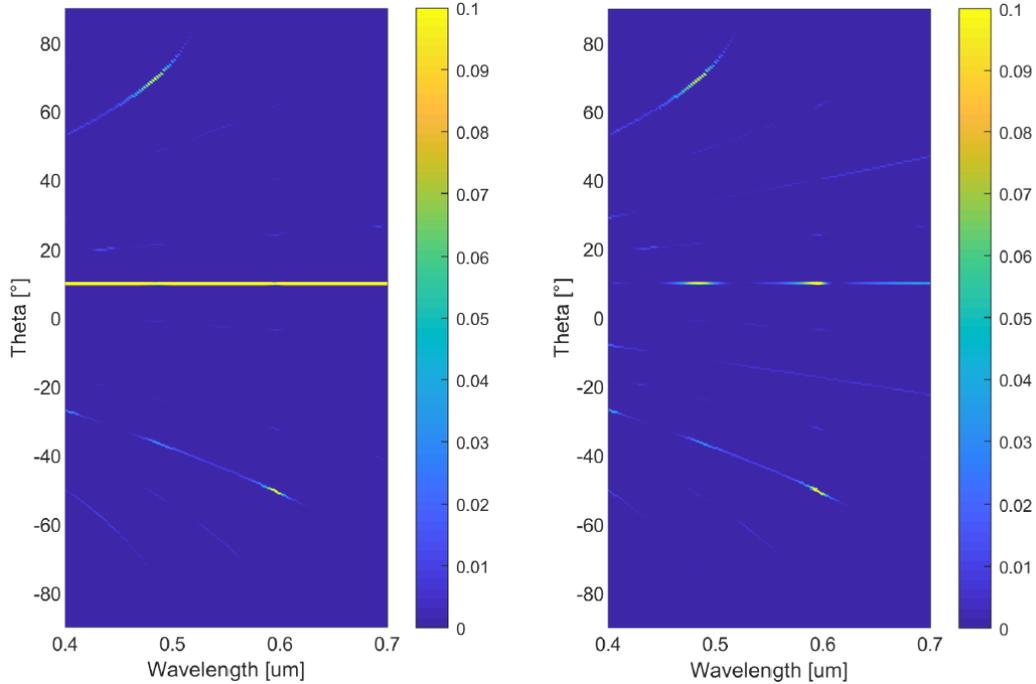

*Figure 28: Light scattering in TE polarization in transmission (left) and reflection (right) for light incident at 10° from normal in collinear incidence, angles in Ormocomp. The -4th and 4th diffraction order can be observed and correspond to guided propagation in the Ormocomp in TIR with respect to an air interface.*

The diffraction of this structure for the TM polarization (not reported here) is much weaker than for TE polarization. This structure can be considered a color selective and polarization specific light coupler, however not a color stable coupler with respect to collinear variations of the light incidence.

More variations are of course possible and other design parameters are not reported here. As a single example the difference of refractive index between the waveguide material and its surroundings – minute refractive index difference are sufficient to provide a strong waveguiding – provides additional flexibility, while asymmetric substrate/superstrate refractive index can provide asymmetric waveguide modes and optical behavior.

## V.  COMPLEX MRWG, THE GENERAL CASE

### i.  Medium-Length Periodic MRWG

Beyond the simple arrangement of small-period MRWGs, having local-perturbations periods that are an integer multiple of their basis period, more complex arrangements allow a large freedom of design. Complex MRWG will be periodic, possibly with large periods, if the basis and the local-perturbations periods are homogeneous, the MRWG periodicity being the least common multiple (LCM) of the two periods, for example expressed in nanometers. However, if the basis period or the LPP are inhomogeneous, for example gradient, the MRWG will be in the general case fully aperiodic. Medium-size period/periodic MRWGs are illustrated with a case similar to the design of part III. A more general design method is provided in part VI.

A regular 200nm period RWG is modified with a 580 LPP made by narrowing the closest individual groove or ridge compared to the LPP locations. This results in a 5.8 microns period MRWG having 10 local perturbations having close to 580nm periodicity.

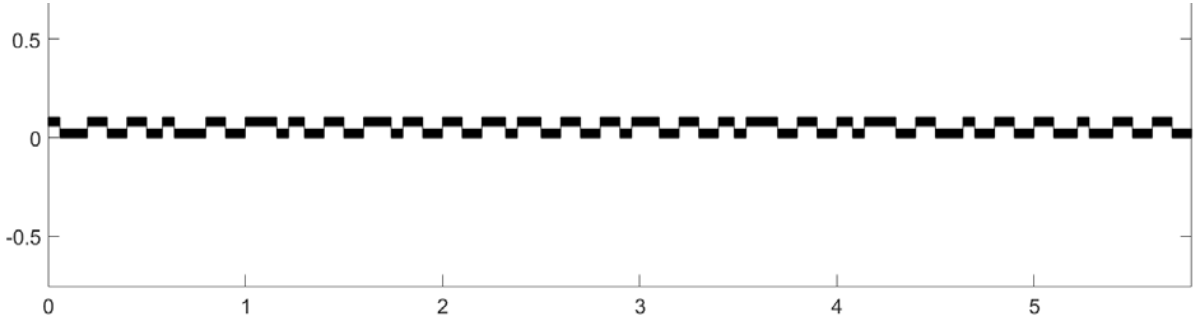

Figure 29: Cross-section schematics of one period of a binary MRWG having a 580nm LPP made by groove or ridge width modulation on a 200nm basis period. The XY scales are in microns. The depth is set at 60nm and the HRI thickness at 45nm. The black rectangles indicate zinc-sulfide domains fully-embedded in a solgel (Ormocomp).

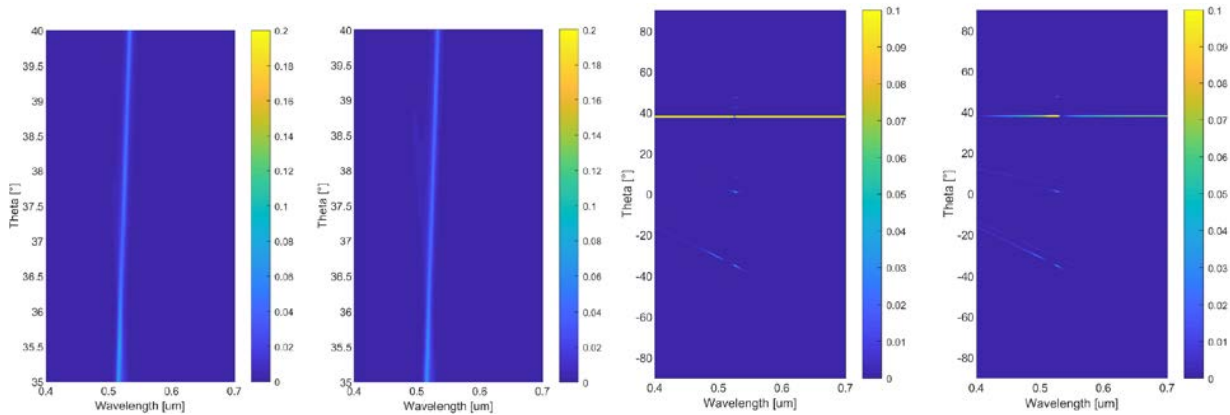

Figure 30: Left: Minus tenth order of diffraction for the 5.8 micron period MRWG described above in TE polarization, with varying incidence angles in Ormocomp. Left most: Transmission, left: Reflection. Right: Scattering in Ormocomp for the 5.8 microns period MRWG at 38° incidence in Ormocomp for TE polarization, Right:: Transmission, Right most: Reflection. In addition to the diffraction order -10, the diffraction order -20 is as well visible as well as traces of other diffraction orders. Such traces can be eliminated with more careful designs and with design relying on longer periodicities.

The HRI thickness tuning allows a spectral tuning of the diffraction band of MRWG using identical replicated nanostructures. Additionally the refractive index of the waveguiding layer can be chosen from the existing dielectric materials which have a good transparency at the wavelength of interest. This allow to modify or not the mode effective index dispersion (refractive index difference can be compensated with the thickness of the waveguiding layer).

ii. An Example of Aperiodic MRWG

As mentioned previously, in many different cases, optical designs will consist in large periodicity MRWG or in only locally periodic or pseudo-periodic gratings which are in fact aperiodic. An example of such an optical function is the realization of an off-axis lens [8] which is color-selective. Such off-axis lens can be used for example as a free space optical combiner [9].

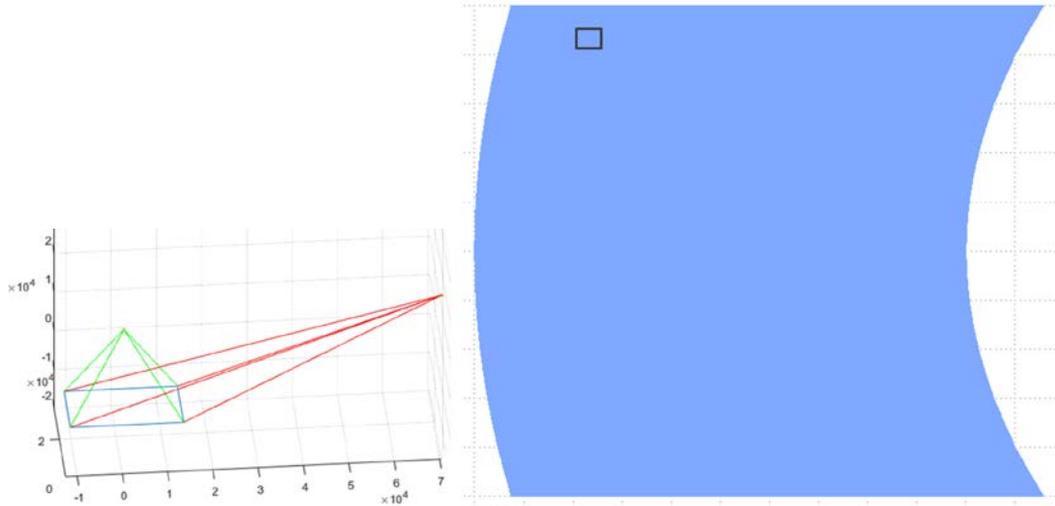

*Figure 31: Geometry and overview of the 2D layout of an MRWG off-axis lens. Left: Geometry of the off axis lens, the green and red segments indicating the chief rays from/to the two focal points and the blue segments indicating the plane of the MRWG. The scale is in microns. Right: Overview of the 2D layout in this MRWG plane, a ROI squared in black is zoomed-in as an examples of a portion of an aperiodic design.*

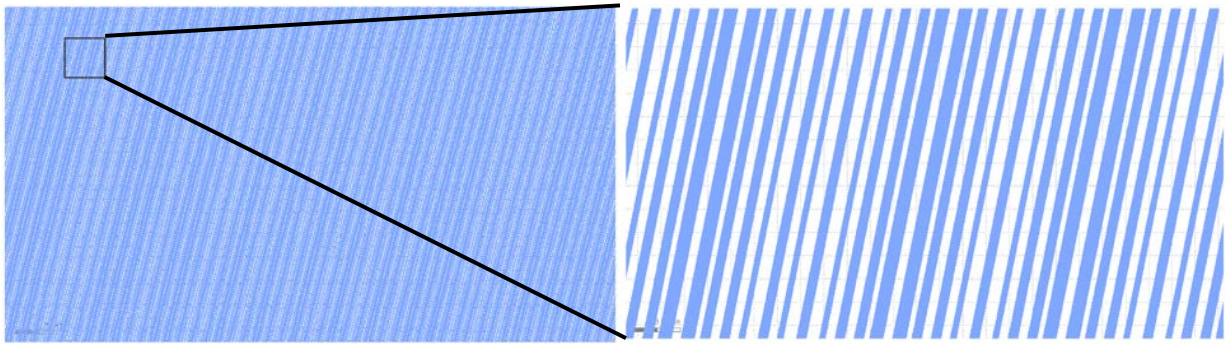

*Figure 32: Successive Zoom-in on the 2D layout of the MRWG off-axis lens shown above from left to right, providing an examples of fully aperiodic binary MRWG designed to operate over a large range of angles with a binary layout and a homogeneous waveguiding layer.*

The manufacturing of the off-axis lens was not fully completed due to the cost of the mastering using e-beam lithography and the characterization of the portion of MRWG realized is not reported here.

## VI. A General Design Flow for MRWG

The various design steps of MRWG are well-known as relying on well-documented optical design principle such as the grating theory, thin waveguide mode computation and numerical modelling methods such as RCWA. An overview how to design such MRWG is provided and consists of a number of different steps, with some parameters of the design impacting various optical properties.

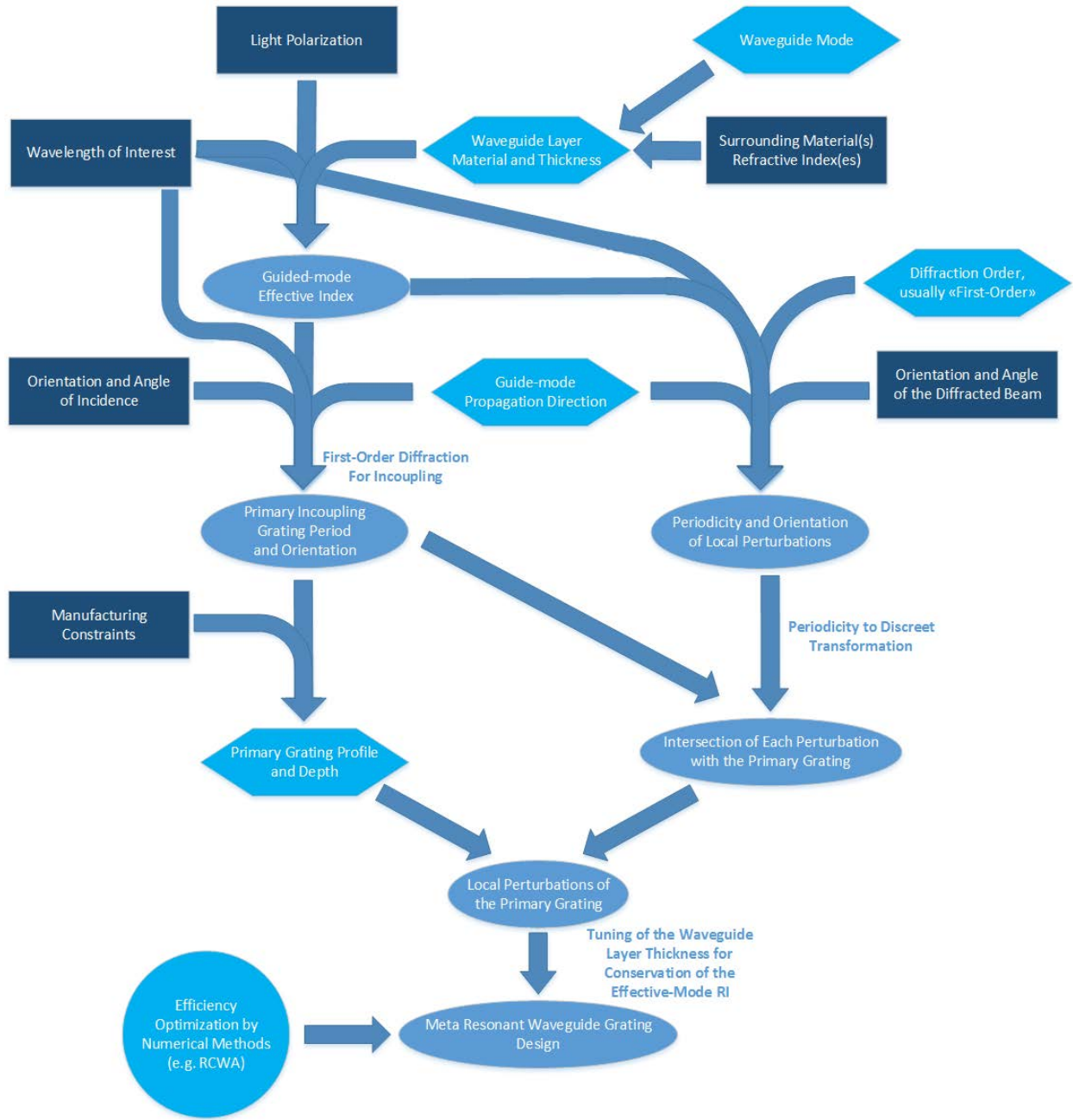

*Figure 33: Typical Design Flow of Meta-Resonant Waveguide-Gratings. Dark-blue rectangles indicate typical design inputs and constraints. Light-blue hexagons are design options or parameters that can be selected/set. Arrows indicate the various design parameters used for the definition or computation of other design parameters, whose results are represented in blue ellipses. Note that the MRWG design is not the final design step as master tools have to be designed with compensation for manufacturing deviations and the master fabrication, for example with e-beam lithography, requires pre-compensation of unwanted discrepancy for various line widths and proximity effects.*

Other design methods using optimization algorithms and neural network such as generative adversarial neural networks can be used. It should be noted that local optimization are not sufficient as the resonances are distributed spatially. This is why, except in specific configurations such as periodic MRWG, optimizations are required on possibly large sets of nanostructures.

# VII. CONCLUSIONS & OUTLOOK

A new type of flat optics is introduced, relying firstly on in-plane light coupling and resonances. Versatile beam steering capabilities with a broad range of possible geometries and large fabrication tolerances is shown. Because of the in-plane light coupling used to interfere with subsequent incidence light, highly wavelength selective and angular selective light redirection is achieved, maintaining a high transparency and clarity to other wavelengths/angles of incidences. As an example, a diffraction bandwidth of less than 10nm is measured with very little remaining diffraction in the whole visible wavelength range, using shallow and easy to replicate structures. Such Meta RWG are intrinsically highly polarization selective because of the different effective index of the TE and TM waveguided mode in the general case, and a complete transparency to one polarization is easily achieved in practice, for example with thin shallow nanostructures diffracting TE-only polarization.

Many different implementations of such Meta Resonant Waveguide Gratings are possible, of which a few are presented using rigorous numerical modelling. This provides a glimpse of the possible design space from a specific standpoint – free-space diffraction in the visible range in TE polarization using a very thin waveguide - with a few experimental results to support the numerical modelling results. However many more cases design options are left to be explored. As example, beyond dielectric thin film waveguiding, implementations relying on long propagation length surface plasmon polariton waves, or on combined dielectric/plasmonic waveguiding can be thought of.

Working with extremely short laser pulses in the femtosecond domains require different analysis to estimate the optical response of these structures, because of the required building of the waveguide mode to perform efficient diffraction. Out of this specific temporal range, the main limitation of the MRWG presented here is the limited diffraction efficiencies, which are much lower than the one achievable with Holographic Optical Elements (HOE). Such limited efficiencies are mostly due to the energy splitting between the zeros and the first diffraction order (with respect to the LPP) in reflection and transmission. Additional work is required to fully understand the theoretical and practical diffraction efficiencies of these systems.

To the opposite of this limitation, MRWG are bringing unique properties, for example complementing HOEs by being able to be more robust and versatile for diffracting a specific frequency range. As examples, MRWG can work at elevated temperatures - when implemented in glasses/inorganic substrates and temperature stable thin films - and operate for a broad set of wavelength ranges - such as in the UV and Medium Infrared Ranges (MIR) - with the existing available transparent materials at these wavelengths, such as silicon thin films in the MIR. Implementation beyond the optical electromagnetic waves are as well most likely possible with the proper materials.

Various applications are not demanding very high diffraction efficiencies and MRWG could be useful due to their ease of manufacturing for visible range MRWG, owing to their compatibility with the many existing roll-to-roll manufacturing lines of DOVIDs. This provides a key advantage with respect with most metasurface concepts, relying on more costly wafer processes with advanced lithography techniques on each wafer.

Beyond complementing existing beam-redirecting optical elements such as HOEs in specific use-cases, other opportunities are expected to rely in the combination of MRWG with other existing optical elements. As an example, MRWG could be combined advantageously with HOE by providing HOE material in close proximity to one of the corrugated waveguide side. This would allow complex and accurate HOE exposure in an holographic setup thanks to the digital design of the MRWG - complex wavefront shaping functions difficult to reach otherwise – and the exposed HOE in optical proximity to MRWG could boost the diffraction efficiency dramatically.

In a different direction, in-plane resonant structures could be combined with phase shifting nanostructures such as vertical waveguide or localized resonator used in dielectric metasurfaces [10, 11], plasmonic metasurfaces, or plasmonic, dielectric and magnetic nanoparticles.

Investigating subwavelength nanostructures arrangements supporting in-plane guided-mode resonances out of the traditional zero-order diffraction could generate new useful optical properties complementing the current well-studied flat diffractive optics with easier to fabricate geometries.

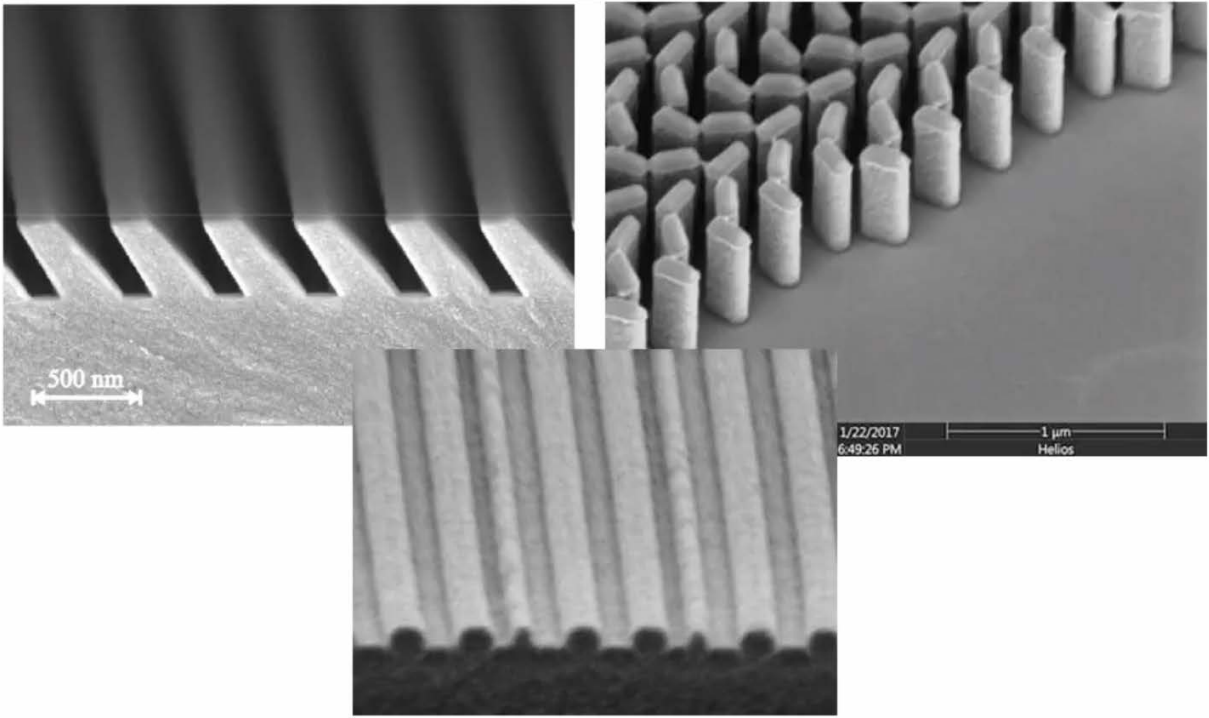

## VIII.   ACKNOWLEDGMENTS

The author would like to thank and acknowledge the work of CEA-LETI, especially Laurent Pain and his team, and Photronics for their support with e-beam mastering, as well as SwissLitho and Finnlitho. Angélique Luu-Dinh and Christian Schneider for their support in sample preparation and SEM imaging. Jérémy Disser and Giorgio Quaranta for support in diffraction and spectroscopy measurements.

# IX.  ANNEX: FROM RWGS TO META-RWGS

## i.  The Development of RWGs

Undesired guided-mode resonances have been first observed by Wood in the reflected diffraction orders of metallic ruled gratings [12]. Rayleigh gave the first valid interpretation of a part of those anomalies five years later in terms of appearance or disappearance of diffracted orders from or into evanescent modes [13]. Only in 1941, Fano proposed that some anomalies may be created by the excitation of surface waves propagating on the grating metallic surface [14]. Hessel and Oliner employed an original theoretical approach based on guided waves rather than on scattering and could explain anomalies of deep grating grooves geometries using numerical tools [15], corroborating the explanation proposed by Fano. Maystre developed a rigorous vector theory in 1972 able to accurately compute the properties of some metallic gratings geometries for any wavelength [16].

In parallel to the discovery of surfaces wave on metallic interfaces, investigations were on-going to use the very high bandwidth provided by light transport in thin film dielectric waveguides in the 1960s. Dakss et al. have experimentally manufactured a grating coupler to thin dielectric films, using the leaky mode in-coupling of a RWG [17]. A few weeks later Kogelnik demonstrated a similar light-coupling using a gelatin volume hologram [18]. In 1973, Nevière, Petit and co-workers developed a rigorous model for the resonances of sinusoidal waveguide-grating couplers in photoresist for transverse electric (TE) polarized light [19], transverse magnetic (TM) polarized light [20] and the computation of the coupling coefficient for finite beams [21] which were confirmed with a high accuracy with experimental data provided by Jacques and Ostrowsky [22].

In addition to in-plane coupling, RWGs have also been extensively investigated for their response in the zeroth order of reflection and transmission. Knop provided a first rigorous model for binary structures such as lamella gratings made of high refractive index dielectrics [3], setting the basis of the rigorous coupled wave analysis (RCWA) modal method; he used this model to compute the zero order resonances and reflective spectra for a large number of more complex geometries of RWGs [23]. Knop proposed furthermore a thin-film coating of a waveguiding layer deposited on a subwavelength-period grating, creating a double-corrugated interfaces geometry, paving the way for the industrial manufacturing of RWG with high-throughput methods. They are nowadays used widely in different application, including optical document security.

## ii.  A definition of RWGs and their main configuration of use

RWGs can have various geometries, for example consisting of high contrast dielectric stripes surrounded by lower refractive index geometries, which are similar to High Contrast Gratings. RWGs are defined here based on their physical behavior, relying on a leaky guided mode propagating over several grating grooves and ridges, rather than on a particular geometry. This definition of RWG based on their physical behavior is necessary due to the continuity between corrugated waveguide geometries and discrete ribbon geometry, as already computed by Knop using rigorous computations in 1981 [23].

A resonant waveguide grating (RWG) consists of a thin waveguiding film in optical contact with a grating. The waveguiding film, by having a higher refractive index than its surrounding media and because of its thin dimension supports a discrete number of guided modes. The waveguide modes can be limited to the fundamental zeroth mode in very thin waveguides or comprise a few modes, in both cases having different mode index for Transverse Electric (TE) and Transverse Magnetic ™ polarizations. Light is coupled into the waveguide modes by the possible grating diffraction orders, depending on the incidence angle and the wavelength. However this guided light is partially diffracted out of the corrugated waveguide at each grating groove and ridge while propagating, providing a leaky-waveguiding. The light scattered out interferes with the non-coupled reflected or transmitted waves providing strong interferences around the coupling wavelength (s). This leads to a very high reflection or transmission of the incident beam depending on the wavelengths and angle of incidence, giving rise to a Fano-like or Lorentzian-like spectral shape at the zeroth order reflection or transmission. Those resonances can have narrow bandwidth, as an example of 0.1 nm FWHM in the visible wavelength range [24]. Depending on the wavelength and phase delay accumulated during propagation in the waveguide, the destructive interference can occur either in reflection or in transmission [25, 26]. Additionally, because the RWG consists usually only of dielectric materials, it can be highly transparent and be used either in transmission or in reflection. RWGs are therefore effective filtering structures, especially for collimated light.

The two main configurations in which RWG have been used - in plane-coupling and zero-order free-space reflection or transmission filtering, and the combination of the two - are due to the low-interest of using RWG in other configurations such as free-space non-zero diffraction. The subwavelength periods of most RWG designs enables only free-space non-zero order diffraction and light-guide coupling for limited wavelength ranges and specific incidence angular ranges. Only a few developments are proposing other configurations.

### iii. Use of RWG in Non Zero-Order or Waveguide Coupling Diffraction

The main developments targeting such other configurations are described. The modification of diffraction gratings with waveguiding layers have been proposed to tune the grating diffraction properties in terms of efficiency and spectral distribution [27] and in terms of transmission versus reflection efficiency [28]. Such configurations are primarily relying on phase delay provided by the different refractive indexed in order to tune the diffraction properties of surface gratings. They are not using a wavelength-selective guided-mode to obtain wavelength specific diffraction but are using the resonant regime of the surface gratings.

Tsitsas investigated the optical behavior of more complex periodic RWG using integral equation analysis [29] without investigating wavelength-selective non-zero diffraction order. Neustock have studied and used multiperiodic RWG [30] for refractive index sensing but only using the zero-order reflection.

Destouches et al. [31] have designed RWGs to be extremely efficient diffraction elements off the Littrow configuration when combined with a dielectric mirror, thus working only in reflection and having very low transmission in the wavelength range of interest.

None of these works could obtain narrowband diffraction of RWG except in zero-order or in plane configurations, to the exception of the work of Destouches with a refection only configuration. Many diffractive optical elements are used in transmission and cannot require comparatively costly to manufacture multilayer dielectric reflectors.

### iv. Pairs of coupled-light exchanging RWGs

Wavelength selectivity, while not possible with conventional diffractive optical elements, is preferable for a large variety of applications. In this direction, Davoine proposed the use of light-exchanging pairs of resonant waveguide gratings (PRWG) [32] to benefit from the wavelength selectivity of RWG in the in-plane coupling configuration, while selecting a different grating period or orientation to outcouple the guided-light to any direction away from direct transmission and specular reflection. These compound diffractive optical elements were designed first to improve refractive index or absorption sensing by avoiding background zero-order reflection and improving the signal to noise ratio (L. Davoine, unpublished PhD work).

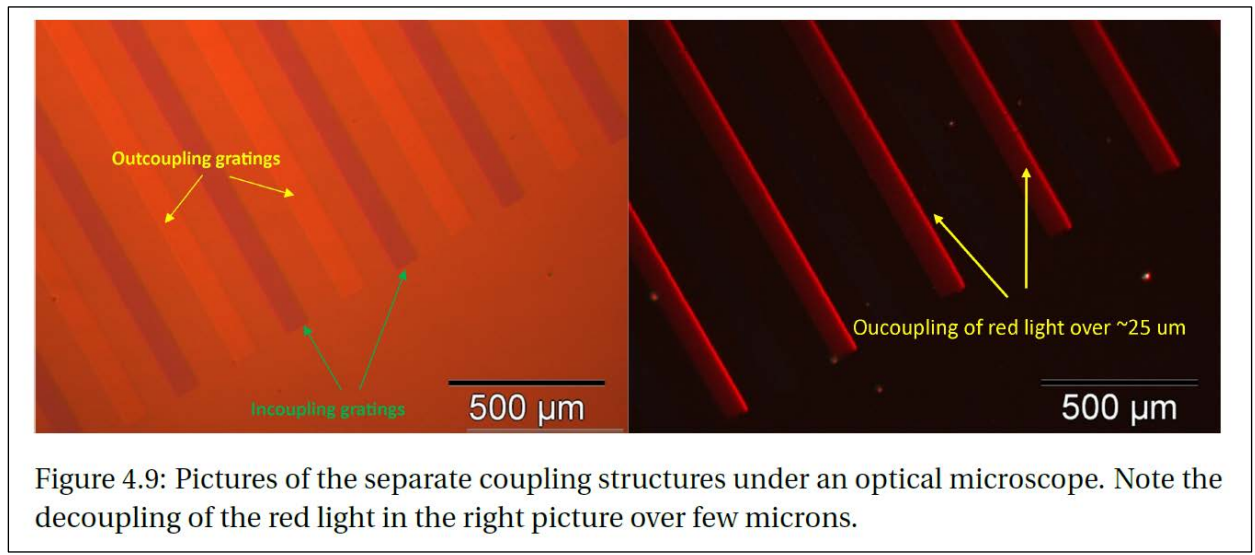

Figure 4.9: Pictures of the separate coupling structures under an optical microscope. Note the decoupling of the red light in the right picture over few microns.

*Figure 34:L. Davoine, unpublished PhD work, 2012*

However they can be useful as well in other applications, such as for opto-digital authentication features for smartphone readout [33].

PRWG are enabling a larger variety of optical behavior than RWG for non-zero-order or in plane configurations. However the necessary discretization and the finite dimensions of the two RWGs in each unit cell are introducing design constraints and limitations, among which: efficiency limitations, wavelength selectivity limitations, lateral beam translations and the intrinsic multispectral-band diffraction. Indeed, up to four diffraction bands are possible out of the zero-order for the simplest cases - fundamental-mode only thin-waveguides - due to the two possible incouplers /outcouplers configuration and the two possible light transport direction - away of normal incidence.

Attempts to mitigate the pixel-related issues introduced by these pairs of RWG by reducing the size of each grating (the number of grating lines) are giving rise to other issues. Very small RWG have poor wavelength selectivity. Additionally arrays of PRWG made of only a few grating lines for each PRWG create additional interference patterns between the various outcoupled beamlets due to the meta-period of the array of PRWG being of only of a few wavelength. A radically different approach is proposed to avoid most of the above listed limitations.

### v. Mixing Two Spatial Periods of a RWG

The optical behavior of optically diffractive structures can be mixed to some extent by superposing the profiles of different diffractive optical elements, which will provide phase shifting spatial maps being the superposition of each diffractive optical elements. As an example, this can be used for the massive superposition of two dimensional reflective phase holograms [34]. An obvious way to combine two surface gratings is by adding their topographies in order to obtain phase delayed being the superposition of the phase delay of the two gratings. If such superpositions are known in reflection and transmission with direct phase shifting, to our knowledge little investigations have been carried out for guided-mode devices. This approach has been used for RWG in zero-order configuration for two gratings having similar periods, providing a large meta-period [35]. This provided two reflection bands balanced by drop of the reflection efficiency.

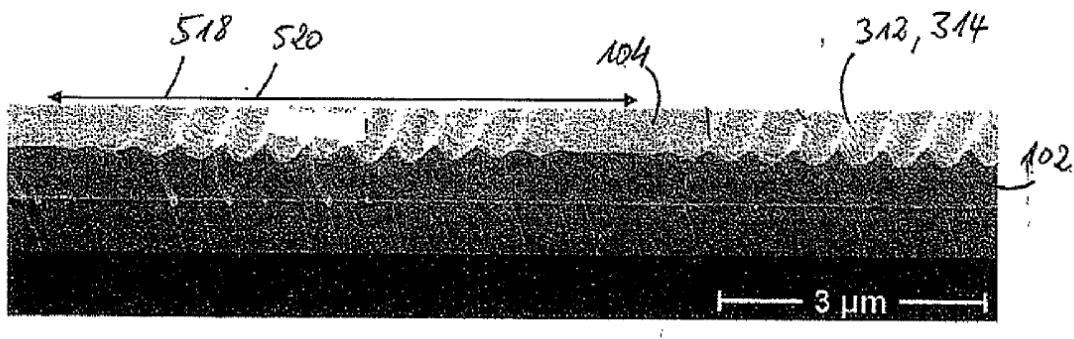

*Figure 35: SEM image of a composite RWG made by the superposition of two periods, reproduced from [27].*